\documentclass[5p, preprint]{elsarticle}
\usepackage{amsmath}
\usepackage{txfonts}
\usepackage{mathrsfs}
\usepackage{graphicx}
\usepackage{dcolumn}
\usepackage{bm}
\usepackage{amssymb}
\usepackage{float}
\usepackage[section]{placeins}
\usepackage{float}
\usepackage[section]{placeins}
\usepackage{multirow}
\usepackage{comment}
\usepackage{color}
\usepackage{slashed}
\usepackage{indentfirst}
\usepackage{epstopdf}
\usepackage{ulem}

\newcommand{\delete}{\bgroup\markoverwith{\textcolor{red}{\rule[0.5ex]{2pt}{1pt}}}\ULon}

\newcommand{\Lrb}[1]{\left\{#1\right\}}
\begin{document}
\title{Superheavy magic structures in the relativistic Hartree-Fock-Bogoliubov approach}
\author[LZU,Orsay]{Jia Jie Li}
\author[LZU]{Wen Hui Long\corref{cor1}}\ead{longwh@lzu.edu.cn}
\author[Lyon]{J\'{e}r\^{o}me Margueron}
\author[Orsay]{Nguyen Van Giai}
\cortext[cor1]{Corresponding author at: School of Nuclear Science and Technology, Lanzhou University, Lanzhou 730000, China}
\address[LZU]{School of Nuclear Science and Technology, Lanzhou University, Lanzhou 730000, China}
\address[Orsay]{Institut de Physique Nucl\'{e}aire, IN2P3-CNRS, Universit\'{e} Paris-Sud, F-91406 Orsay Cedex, France}
\address[Lyon]{Institut de Physique Nucl\'{e}aire de Lyon, IN2P3-CNRS, Universit\'{e} de Lyon, F-69622 Villeurbanne Cedex, France}
%
%
\begin{abstract}
We have explored the occurrence of the spherical shell closures for superheavy nuclei in the framework of the relativistic
Hartree-Fock-Bogoliubov (RHFB) theory.
Shell effects are characterized in terms of two-nucleon gaps $\delta_{2n(p)}$. Although the results depend slightly on
the effective Lagrangians used, the general set of magic numbers beyond $^{208}$Pb are predicted
to be $Z = 120$, $138$ for protons and $N = 172$, 184, 228 and 258 for neutrons, respectively.
Specifically the RHFB calculations favor the nuclide $^{304}$120 as the next spherical doubly magic one beyond $^{208}$Pb.
Shell effects are sensitive to various terms of the mean-field, such as the spin-orbit coupling, the scalar and effective
masses. 
\end{abstract}
\maketitle
%
%
%
%

For a fairly long period, it remains a challenging issue in nuclear physics to explore the existence
limit of very heavy nuclei, i.e., the superheavy elements (SHE) with $Z \geq 104$ and the so-called
stability island of superheavy nuclei (SHN). 
If at all, the existence of this island in the nuclear chart would come from very subtle
contributions to the nuclear binding energy~\cite{Block2010}.
Experimentally, the discoveries of new elements up to $Z = 118$ have been reported in
Refs.~\cite{Oganessian2006, Oganessian2010}.
The increasing survival probabilities with increasing proton number of SHE from $Z = 114$
to $118$ seem to indicate enhanced shell effects with increasing $Z$ and therefore a possible proton
magic shell may emerge beyond $Z \geq 120$~\cite{Adamian2009}.

On the other hand, theoretical studies have provided a large amount of valuable information for the
exploration of SHN.
These studies can be separated into different categories: Microscopic - Macroscopic (Mic-Mac)
models~\cite{Moller1995, Baran2005}, non-relativistic mean field~\cite{Rutz1997, Bender1999, Decharge2003}
and covariant mean field~\cite{Rutz1997, Bender1999, Zhang2005} approaches.
The extrapolation towards the superheavy region challenges the predictivity of nuclear models.
The Mic-Mac approach, despite its great success in predicting nuclear binding energies for exotic nuclei,
can hardly be extrapolated towards very new regions where experimental data are extremely scarce.
The stability of nuclei is mostly driven by shell effects and therefore, self-consistent mean field methods are
probably the best conceptual tool to explore the superheavy region, although the Mic-Mac models still
give a better quantitative description of heavy nuclides.

We are searching for doubly closed-shell systems and we assume spherical symmetry.
Then, the shells are essentially determined by the spin-orbit (SO) splittings, and by the effective masses. 
Another effect which affects the shell structure is related to the possible occurrence 
of an almost degeneracy among pseudo-spin (PS) partners~\cite{Ginocchio2005, Long2007}.
In the non-relativistic self-consistent mean field theory~\cite{Bender2003, Stone2007}, the SO splittings depend directly on an extra SO parameter in the energy density functional.
In the superfluid covariant density functional (CDF) theory, like the relativistic
Hartree-Bogoliubov (RHB)~\cite{Vretenara2005, Meng2006} or the relativistic Hartree-Fock-Bogoliubov (RHFB)~\cite{Long2010a} approaches, the SO splitting
depends directly on the Lorentz scalar and vector mean fields without additional term.
The SO splitting is not adjusted and can be considered as a prediction of relativistic
Lagrangians, even in ordinary nuclei.
This might be an advantage for exploring unknown regions.
Furthermore, in the more complete RHFB version of the CDF theory the SO splittings
can be affected by meson-nucleon couplings like Lorentz $\rho$-tensor couplings~\cite{Long2007}
not present in the simple RHB.
This is one of the main motivations for undertaking the present study in the framework of the RHFB
approach.

In this work we investigate the superheavy nuclides covering $Z = 110-140$.
In the pairing channel, the finite-range Gogny force D1S~\cite{Berger1984} renormalized by a
strength factor $f$ is adopted as the effective pairing interaction.
The strength factor $f$ is introduced to compensate level-density differences among various
mean field approaches.
It was indeed shown that pairing related quantities, such as odd-even mass differences and moments of
inertia, are systematically overestimated in the RHFB calculations of heavy nuclei with
the original Gogny pairing force~\cite{Wang2013}. 
The strength factor $f=0.9$ is therefore 
adjusted to reproduce 
the odd-even mass differences of odd Pb isotopes.
Concerning the relativistic Hartree-Fock (RHF) mean field, the adopted effective Lagrangians are PKA1~\cite{Long2007} and
the PKOi series (i=2, 3)~\cite{Long2006a, Long2008}.
To compare with approaches neglecting the Fock term (RHB), we also
use PKDD~\cite{Long2004} and DD-ME2~\cite{Lalazissis2005} Lagrangians.
The integro-differential RHFB equations are solved 
by using a Dirac Woods-Saxon
basis~\cite{Zhou2003} with a radial cutoff $R = 28$~fm.
The numbers of positive and negative energy states in the basis expansion for each single-particle (s.p.)
angular momentum ($l, j$) are chosen to be 44 and 12, respectively.
%
%
%
%

\begin{figure}[t]
\centering
\ifpdf
\includegraphics[width = 0.47\textwidth]{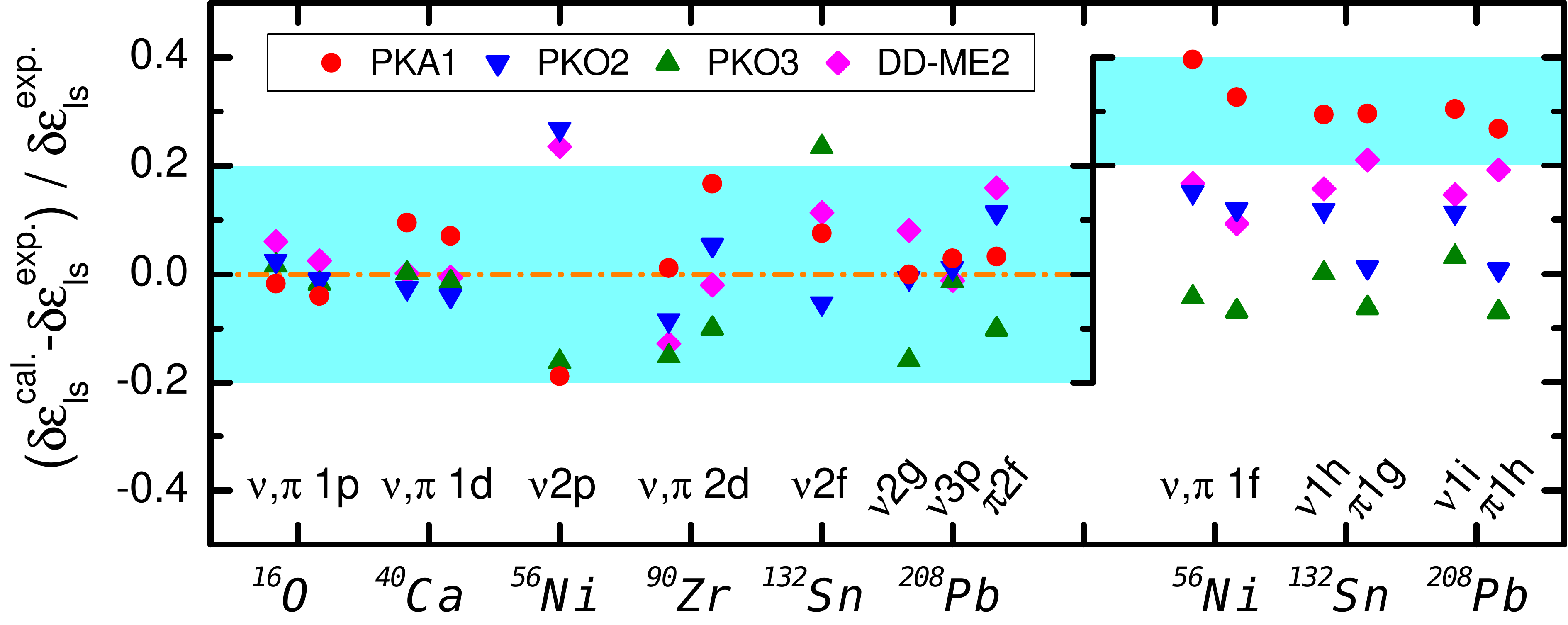}
\else
\includegraphics[width = 0.47\textwidth]{SO.eps}
\fi
\caption{
Relative differences between the theoretical SO splittings $\delta\epsilon_{ls}^{cal.}$
and the experimental ones $\delta\epsilon_{ls}^{exp.}$~\cite{Nudat2013} in the (semi)-doubly magic nuclei indicated
on the horizontal-axis. Particle and hole SO partners are shown on the left while particle-hole ones are on the right. See the text for details.}
\label{fig:SO}
\end{figure}

Let us first discuss extrapolations to SHE of mean field models which are well constrained on medium and heavy nuclei. For instance, due to the high level density in SHE, small variations in the s.p. level spacings due to different 
SO splitting predictions of various models, can have a large effect on
magicity. The SO force is therefore a crucial ingredient of nuclear structure models, especially when it
comes to extrapolations to SHN.
Fig.~\ref{fig:SO} shows the 
relative differences between calculated and experimental~\cite{Nudat2013} SO splittings for
a selection of  
levels having well controlled spectroscopic factors.
The relative differences are typically $\sim 20\%$ when both partners are particle or hole states, but they become larger otherwise. This is not surprising since 
polarization and correlation effects tend to shift unoccupied and occupied s.p. states into opposite directions ~\cite{Rutz1998, Litvinova2006}. 
If one compares the results of Fig.~\ref{fig:SO} with those from non-relativistic mean field models such as Skyrme-Hartree-Fock (SHF)~\cite{Bender1999} 
it appears that the latter give systematically larger deviations. Fig.~\ref{fig:SO} provides therefore a good motivation for predictions of SHE based on relativistic
Lagrangians.

SHE predictions have been carried out using relativistic mean field (RMF)
models~\cite{Rutz1997} or RHB models~\cite{Zhang2005}.
In such Hartree-type approaches, the contribution of the Fock term is disregarded, at variance with
RHF, leading to a renormalization of the coupling constants.
It is an approximation which forbids the inclusion of the $\pi$ and the $\rho$-tensor mesons.
While RMF models are as predictive as RHF ones for medium and heavy nuclei, it is preferable
to base extrapolations to SHE on calculations including correctly the contribution of the Fock term.
It is also a motivation of the present study.

Magicity in SHN might not be as well-marked as in the ordinary nuclei~\cite{Bender1999}.
To identify the magic shells, we will employ the so-called two-nucleon gaps, $\delta_{2p}$ (proton) and $\delta_{2n}$ (neutron), i.e., the difference of
two-nucleon separation energies of neighboring isotopes or isotones, which provides an efficient
evaluation of the shell effects~\cite{Rutz1997, Zhang2005},
\begin{subequations}\label{nucleon gaps}
\begin{eqnarray}
\delta_{2p}(N,Z)&=&S_{2p}(N,Z)-S_{2p}(N,Z+2),\\
\delta_{2n}(N,Z)&=&S_{2n}(N,Z)-S_{2n}(N+2,Z).
\end{eqnarray}
\end{subequations}
The peak values of the two-nucleon gaps are essentially determined by the sudden jump of the two-nucleon separation energies, which can be taken as a clear evidence of the magic shell occurrence.

\begin{figure*}[t]
\ifpdf
\includegraphics[width = 1.00\textwidth]{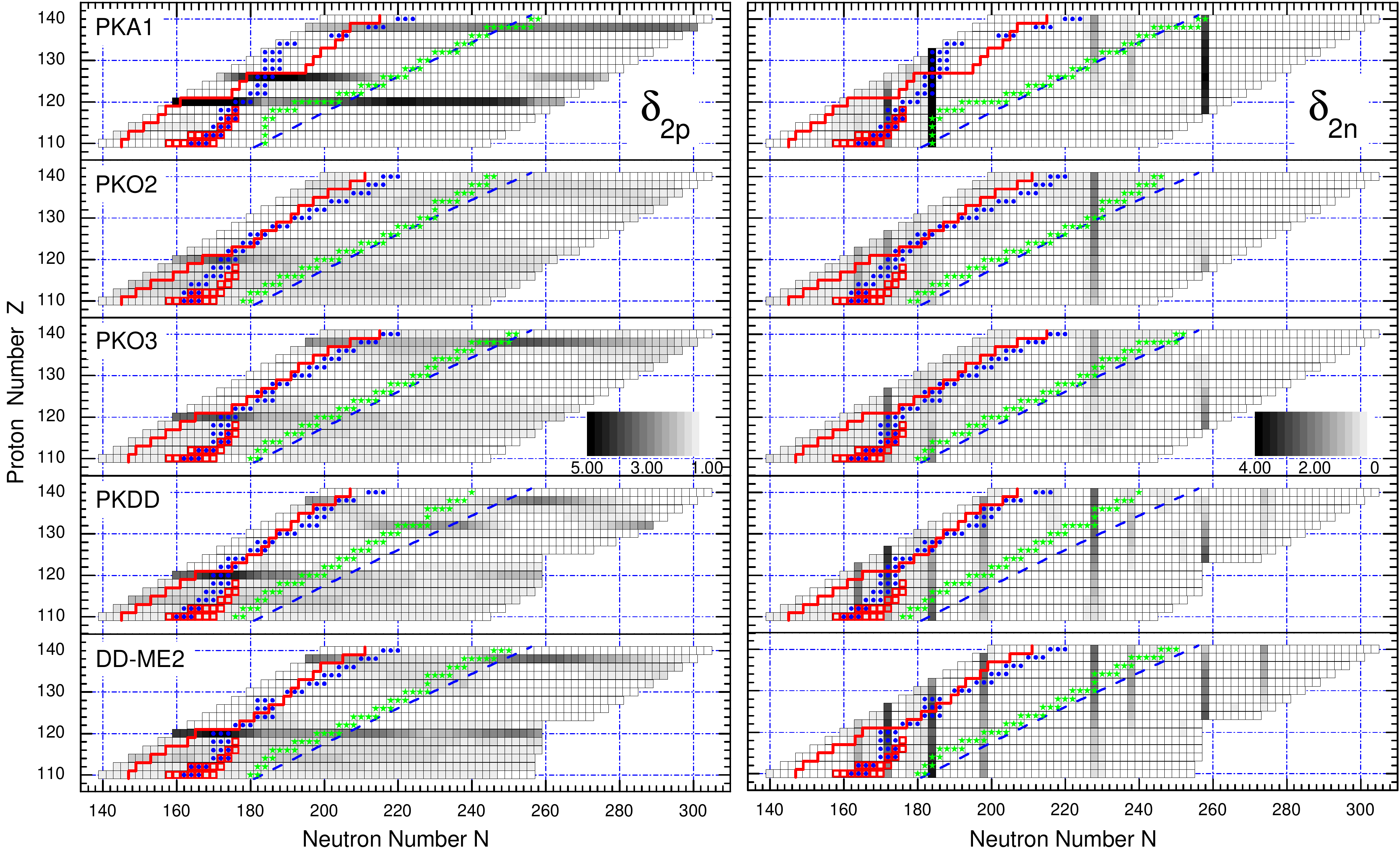}
\else
\includegraphics[width = 1.00\textwidth]{D2N.eps}
\fi
\caption{Contour plots (in MeV) for the two-proton gaps $\delta_{2p}$ (left panels) and the two-neutron gaps $\delta_{2n}$
(right panel) as functions of N and Z. The two-nucleon gaps are obtained with PKA1, PKO2 and PKO3 parametrizations for RHFB, and PKDD and DD-ME2 parametrizations for RHB. 
The red-solid lines represent the two-proton drip lines. Nuclei stable with respect to $\beta$-decay and fission are marked with green filled stars and blue circles, respectively. The blue long-dashed lines represent the empirical $\beta$-stability line~\cite{Marmier1971}.
The red empty squares indicate the experimental SHN from NUSEBASE2012~\cite{Audi2013}. See text for more details.}
\label{fig:DN}
\end{figure*}

Fig. \ref{fig:DN} presents the two-proton (left panels) and two-neutron (right panels) gaps for the $Z = 110-140$ 
even-even isotopes calculated with the selected effective Lagrangians. We have adopted the presentation of Ref.~\cite{Rutz1997}
so that the similarities and differences in the predictions of the earlier study can be more easily seen.
The red-solid lines stand for the two-proton drip lines defined as the change in sign of the two-proton separation energy.
Nuclei that are stable with respect to $\beta$-decay or fission are represented with filled green stars or filled blue
circles, respectively.
For a given $A$ (resp. $Z$), the $\beta$-stability (resp. fission-stability) line is located at the
maximum of the binding energy per nucleon, and corresponds as well to the minimum of the $Q$-value
for $\beta$-decay (resp. fission)~\cite{Wu1996}. The dashed blue line represents the $\beta$-stability line
given by the empirical formula $Z = A/(1.98+0.0155A^{2/3})$~\cite{Marmier1971}.
Experimental data taken from the NUBASE2012 evaluation of nuclear properties~\cite{Audi2013}, 
including the extrapolated SHN, are located below $Z = 118$ and are shown in Fig.~\ref{fig:DN} with empty red squares. 
It is observed from Fig.~\ref{fig:DN} that these nuclei coincide largely with the nuclei which are stable
with respect to fission (filled blue circles), as predicted by our models, especially by PKA1.
The effects of deformation are not included in the present study of $\delta_{2p}$ and $\delta_{2n}$ 
although they may also play a significant role~\cite{Lalazissis1996, Patra1999}.
On the neutron-poor side, the large Coulomb barrier existing in SHN pushes further down the
two-proton drip line. The effect is expected to change by a few units the position of the drip line.

In Fig.~\ref{fig:DN}, the squares are filled in proportion of the gap, which varies from 1 to 5~MeV,
as shown in the grey-scale index. Structures with large gaps between 3 and 5 MeV appear clearly in Fig.~\ref{fig:DN}.
From the comparison of the different models shown in Fig.~\ref{fig:DN}, it is clear that PKA1 is the
Lagrangian which predicts the larger gaps for $Z = 120$, 126, 138 and $N = 184$, 258. These numbers
are thus the predicted magic numbers in neutron-rich SHN based on the PKA1-RHFB model.
The other effective Lagrangians also present a remarkable proton shell at $Z=120$. In addition,
$Z = 132$ for PKDD-RHB and $Z = 138$ for both RHFB (PKA1 and PKOi) and RHB (PKDD and DD-ME2) approaches 
are found to be possible proton magic numbers, consistent with the predictions in
Ref.~\cite{Zhang2005}. Concerning the neutron shells, besides $N=184$ and 258, PKA1 also presents a well-marked shell structure at $N = 172$, which is also present in the predictions of the other Lagrangians.
Fairly distinct shell effects at $N = 184$ and 258 are also found with the other parameterizations, except
with PKO2. Remarkable shell effects are found at $N = 228$, although less pronounced compared to those 
at $N = 184$ and $258$ predicted by PKA1. Furthermore, a neutron shell is predicted at $N = 164$ with PKO2, PKDD and DD-ME2 models, and another is predicted at $N = 198$ with RHB models (PKDD and DD-ME2).

We have checked that the neutron and proton pairing gaps are also quenched for the same proton
and neutron magic numbers as those obtained in Fig.~\ref{fig:DN} for each considered Lagrangian.
Combined with the two-nucleon gaps, it is found that the proton shell $Z = 120$ is predicted by PKA1 
as well as by the other Lagrangians used in Fig.~\ref{fig:DN}.
It is also predicted by some SHF models such as SLy6, SkI1, SkI3 and SkI4~\cite{Rutz1997}, but it must be 
stressed that the SHF models can give different predictions for $Z = 114$ and $Z = 126$, see for instance Ref.~\cite{Rutz1997}.
$Z=120$ can however be considered as a fairly good candidate for proton magic number. 
In Ref.~\cite{Rutz1997} SHF forces such as SkM* or SkP predict $Z = 126$ as a magic number
for neutron-poor isotopes. 
$Z = 126$ is also predicted as a magic number by PKA1 model, but not by the other Lagrangians 
used in Fig.~\ref{fig:DN}, which predict a weak SO splitting for 
high-$j$ states.
  
On the other hand, the situation for the neutrons is more complex. 
Although $N = 172$ and 228 magic numbers seem to
be generally predicted by the selected effective Lagrangians, the corresponding
shell effects are rather weak.
Except for PKO2, $N = 184$ and $258$ are also generally predicted as candidates for neutron magic numbers.
Let us notice that a large number of SHF models considered in Ref.~\cite{Rutz1997} as well as Gogny forces~\cite{Decharge2003} have also a large gap for these neutron numbers.
Specifically, PKA1 can provide a better description of the nuclear shell structure than the others~\cite{Long2007}
and a better agreement on the fission stability of observed SHN (see Fig.~\ref{fig:DN}), and it leads to  
pronounced shell effects. In fact, as indicated by SHF investigations~\cite{Kruppa2000} $N = 184$ is also
favored evidently to be a spherical neutron magic number and the $N = 184$ isotones are expected to
have spherical shapes. By comparing the predictions between the various models discussed here, we conclude that $^{304}120_{184}$ 
is a most probable doubly magic system in the SHN region, and $^{292}120_{172}$ might be another candidate with 
less stability.
%
%
\begin{table}[tb]
\caption{Bulk properties of symmetric nuclear matter calculated with the effective interactions PKA1, PKOi series, PKDD and DD-ME2: saturation density $\rho_0$ (fm$^{-3}$), binding energy per particle $E_{B}/A$ (MeV), incompressibility $K$ (MeV), asymmetry energy coefficient $J$ (MeV), scalar mass $M^\ast_S$ and non-relativistic
effective mass $M^\ast_{NR}$ in units of nucleon mass $M$.}\setlength{\tabcolsep}{5pt}
\label{tab:NMP}
\begin{tabular}{lcccccc}
\hline\hline
 Force  &$\rho_0$ & $E_{B}/A$ & $K$ & $J$ & $M^\ast_S$ & $M^\ast_{NR}$ \\
\hline
 PKA1   &  0.160 & $-15.83$ &  229.96 &  36.02 &  0.547  & 0.681  \\
 PKO1   &  0.152 & $-16.00$ &  250.28 &  34.37 &  0.590  & 0.746  \\
 PKO2   &  0.151 & $-16.03$ &  249.53 &  32.49 &  0.603  & 0.764  \\
 PKO3   &  0.153 & $-16.04$ &  262.44 &  32.98 &  0.586  & 0.742  \\ \hline
 PKDD   &  0.150 & $-16.27$ &  262.18 &  36.79 &  0.571  & 0.651  \\
 DD-ME2 &  0.152 & $-16.14$ &  250.97 &  32.31 &  0.572  & 0.652  \\
\hline\hline
\end{tabular}
\end{table}

Nevertheless, from Fig.~\ref{fig:DN} one can find distinct deviations among the models in predicting the magic numbers. $Z = 120$ can be considered as a reliable prediction of proton magic number and $Z = 138$ could be another candidate with more model dependence. The neutron shells $N = 172$, 184, 228 and 258 are
common to several models. Other shells, e.g., $N = 198$, appear essentially model dependent. Among the present results, one may notice that RHB calculations (PKDD and DD-ME2) predict more shell closures than RHFB, and PKO2-RHFB predicts fewest candidates. To interpret such distinct deviations, Table~\ref{tab:NMP} shows the bulk properties of symmetric nuclear matter determined by the present sets of Lagrangians.
In general the occurrence of superheavy magic shells is closely related with both the scalar mass $M^\ast_S$ and effective mass $M^\ast_{NR}$~\cite{Long2006a}, which essentially determine the strength of SO couplings and level densities, respectively. Among the present models, the effective Lagrangian PKO2 predicts the largest values of both masses, leading to relatively weak SO couplings and high level density on the average. As a result there remains little space in the spectra for the occurrence of magic shells. On the other hand, the RHB models (PKDD and DD-ME2) predict more magic shells due to the relatively small masses. In fact, as seen from Fig.~\ref{fig:DN}, PKO2 also presents weaker shell effects than the others. For PKA1 the situation is different. Although it has a larger effective mass $M^\ast_{NR}$ than PKDD or DD-ME2, PKA1 gives a smaller scalar mass $M_S^*$ and shows stronger shell effects than the others. These may partially explain why PKA1 does not suffer from the common drawback of the CDF calculations --- the so-called artificial shell closures induced by low  $M_S^*$ and $M^\ast_{NR}$~\cite{Geng2006} --- and why 
it leads to more degenerate PS partners~\cite{Long2007, Long2009, Long2010b}. 
%
%
%
%
\begin{figure*}[tb]\centering
\ifpdf
\includegraphics[width = 0.70\textwidth]{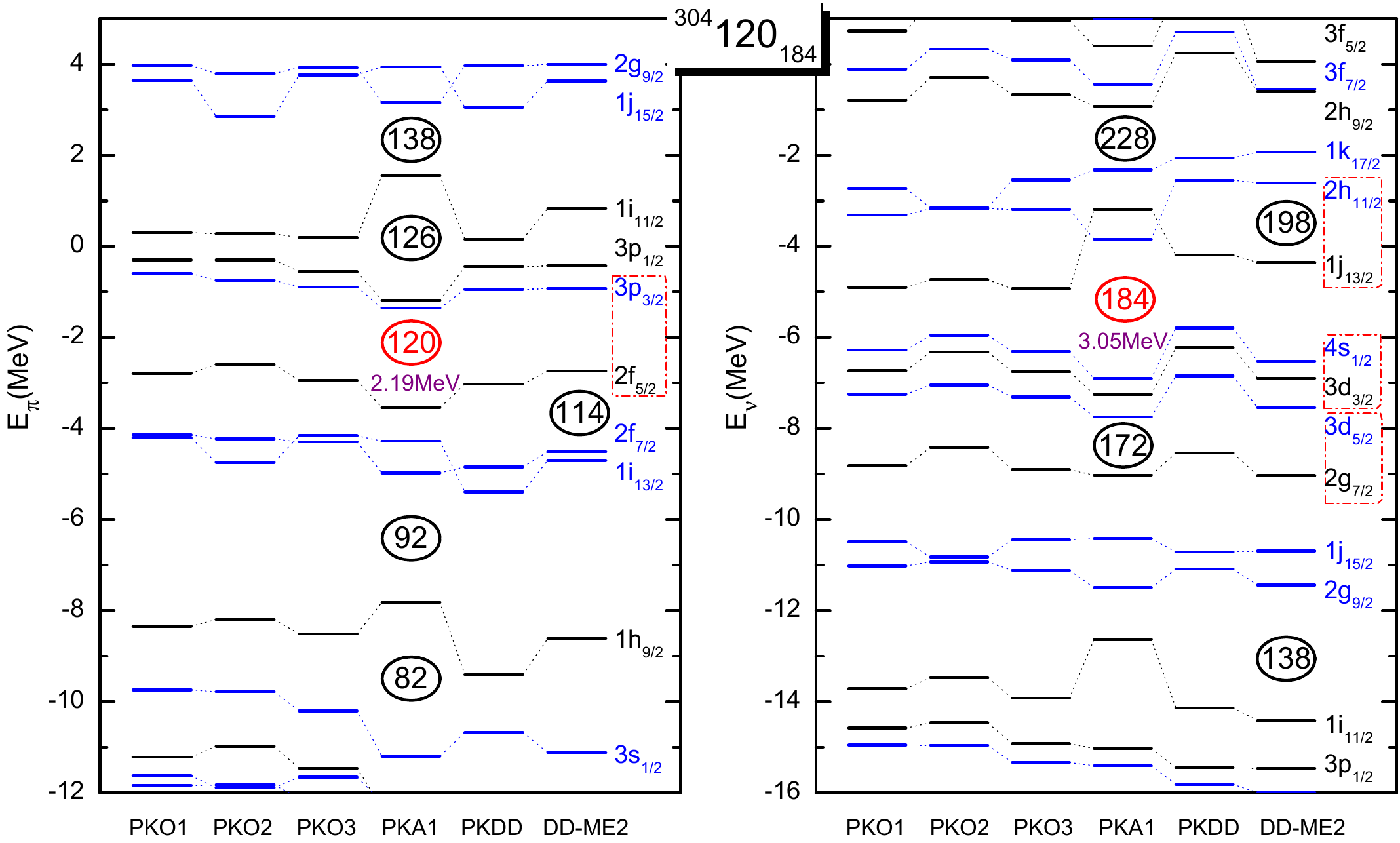}
\else
\includegraphics[width = 0.70\textwidth]{SPS120184.eps}
\fi
\caption{Proton (left panel) and neutron (right panel) canonical s.p. spectra of superheavy nuclide $^{304}120$. The results are extracted from the RHFB calculations with PKOi series and PKA1, and compared to the RHB ones with PKDD and DD-ME2. In all cases the pairing force is derived from the finite range Gogny force D1S with the strength factor $f = 0.9$. See the text for details.}
\label{fig:SPS}
\end{figure*}

It is interesting to compare the s.p. spectra obtained with different effective Lagrangians.
Taking the doubly magic SHN $^{304}120_{184}$ as an example, Fig.~\ref{fig:SPS} shows the proton
(left panel) and neutron (right panel) canonical s.p. spectra provided by selected models.
It is found that PKA1 provides the most evident magicity at $Z = 120$ and $N = 184$, respectively,
although these shell closures are much weaker than in ordinary nuclei.
For the neutron shell $N = 184$, it is essentially determined by the degeneracy of two 
PS partners $\Lrb{2h_{11/2}, 1j_{13/2}}$ and $\Lrb{4s_{1/2}, 3d_{3/2}}$, respectively above and below the shell.
For the latter, the PS partners are predicted to be almost 
degenerate by all the models considered, while for the former, the PS partners have high angular momentum and some 
differences among models are observed: PKA1 predicts a weak PS splitting, at variance
with the predictions of the other Lagrangians.

It is interesting to discuss the structure of the s.p. levels for the proton shell closure $Z=120$.
As shown in the left panel of Fig.~\ref{fig:SPS} the proton shell closure coincides with a
large PS splitting, $\Lrb{3p_{3/2}, 2f_{5/2}}$, whereas the SO doublet $\Lrb{3p_{1/2}, 3p_{3/2}}$
above the shell is almost degenerate.
The shell gap at $Z = 120$ can therefore be interpreted as a manifestation of 
a large PS splitting and a weak SO splitting.
Below the shell $Z = 120$, the protons filling in the high-$j$ states will be driven towards the surface
of the nucleus due to the strong centrifugal potential and large repulsive Coulomb field in SHN.
Both effects lead to an interior depression of the proton distributions and consequently the interior region of the mean potential is not flat any more~\cite{Decharge2003}. 
As a result the SO splitting is reduced, particularly for the low-$l$ states $3p$ and $2f$ which have more overlap with the interior depression. Consequently the splitting between neighboring PS partners (i.e., $3p_{3/2}$ and $2f_{5/2}$) is somewhat enlarged~\cite{Shen2013}. 
In Ref.~\cite{Afanasjev2005} it is also pointed out that the pronounced central depressions in the densities lead to the spherical 
shell gaps at $Z = 120$ and $N = 172$ as a direct consequence of 
a large PS splitting, whereas a flatter density profile favors the shell occurrence at $N = 184$ 
and $Z = 126$. This can happen not only for SHN, and the emergence of a new shell closure at $Z$ or $N = 16$ and $N = 32$~\cite{Ozawa2000, Kanungo2002} can be also related 
to a similar mechanism in light exotic nuclei.
%
%
%

In summary, we have explored the occurrence of spherical shell closures for SHN and the physics therein using the RHFB theory with density-dependent meson-nucleon couplings, and compared the predictions with those of some RHB models. The shell effects are quantified in terms of two-nucleon gaps $\delta_{2n(p)}$. 
To our knowledge, this is the first attempt to perform such extensive calculations within the RHFB scheme. The results indicate that the nuclide $^{304}120_{184}$ could be the next spherically doubly magic nuclide beyond $^{208}$Pb. It is also found that the shell effects in SHN are sensitive to the values of both scalar mass and effective mass, which essentially determine the spin-orbit effects and level density, respectively.
As we already pointed out in the introduction, the emergence or disappearance of shell closure is tied up with the evolution of the central and spin-orbit mean fields, 
a feature that covariant mean field models may describe in a more unified way as compared to non-relativistic energy density functional (EDF) approaches. A further advantage of the RHFB framework 
is that exchange (Fock) terms are explicitly treated rather than approximately included by readjusted direct (Hartree) contributions as it is done in RHB (this is particularly true 
for the Coulomb exchange energy which is basically absent in RHB).   

Experimental measurement of $Q_\alpha$ for at least one isotope of $Z = 120$ nucleus would help to set a proper constraint in determining the shell effects of SHN and to test further the reliability of the models as well. One also has to admit that for a more extensive exploration one needs to take into account the deformation effects.
%
%
%

We would like to thank Haozhao Liang for enlightening discussions on PS splitting. We thank the referee for his helpful and encouraging remarks. 
One of us (J. Li) also thanks Haifei Zhang for his help in the initial stage of this study. This work is partly supported by the National Natural Science Foundation of China under Grant No. 11075066, the Fundamental Research Funds for the Central Universities under Contracts No. lzujbky-2012-k07, and the Program for New Century Excellent Talents in University under Grant No. NCET-10-0466, and the Specialized Research Fund for the Doctral Program of Higher Education under Grant No. 20130211110005.
%
%


\end{document}